\documentclass[preprint,showpacs,preprintnumbers,amsmath,amssymb]{revtex4}
\usepackage{amssymb}
\usepackage{amsmath}
\usepackage{graphicx}
\usepackage{epsfig}
\usepackage{subfigure}
\usepackage{amsfonts}
\usepackage{CJK}

\begin{document}

\title{Generation of quantum entangled states of multiple groups of qubits distributed in multiple cavities}

\author{Tong Liu$^{1}$}
\author{Qi-Ping Su$^{2}$}
\author{Yu Zhang$^{3}$}
\author{Yu-Liang Fang$^{1}$}
\author{Chui-Ping Yang$^{1}$}
\email{yangcp@hznu.edu.cn}

\address{$^1$Quantum Information Research Center, Shangrao Normal University, Shangrao 334001, China}
\address{$^2$Department of Physics, Hangzhou Normal University, Hangzhou 311121, China }
\address{$^3$School of Physics, Nanjing University, Nanjing 210093, China}

\date{\today}

\begin{abstract}
Provided that cavities are initially in a Greenberger-Horne-Zeilinger (GHZ) entangled state, we show that GHZ states of $N$-group qubits distributed in $N$ cavities can be created via a 3-step operation.
The GHZ states of the $N$-group qubits are generated by using $N$-group qutrits placed in the $N$ cavities. Here, ``qutrit" refers to a
three-level quantum system with the two lowest levels representing a qubit while the third level acting as an intermediate state necessary for the GHZ state creation. This proposal does not depend on the architecture of the cavity-based quantum network and the way for coupling the cavities. The operation time is independent of the number of qubits. The GHZ states are prepared deterministically because no measurement on
the states of qutrits or cavities is needed. In addition, the third energy level of the qutrits during the entire operation
is virtually excited and thus decoherence from higher energy
levels is greatly suppressed. This proposal is quite general and can in principle be applied to create GHZ
states of many qubits using different types of physical qutrits (e.g., atoms, quantum dots, NV centers, various superconducting qutrits, etc.)
distributed in multiple cavities. As a specific example, we further discuss the experimental feasibility of preparing a
GHZ state of four-group transmon qubits (each group consisting of three qubits) distributed in four one-dimensional transmission line resonators arranged in an array.
\end{abstract}
\maketitle
\date{\today }

\begin{center}
\textbf{I. INTRODUCTION AND MOTIVATION}
\end{center}

Large-scale quantum information processing (QIP) has drawn much attention [1-3].
Usually, a large number of qubits may be involved in large-scale QIP. The size of QIP with
qubits in multiple cavities can be larger when compared to QIP with qubits
in a single cavity. For instance, given the number of qubits in each cavity
is $m$, the number of qubits placed in $n$ cavities is $n\times m$, which is
$n$ times the number $m$ of qubits placed in a single cavity. Therefore,
large-scale QIP based on cavity or circuit QED may require distributing
qubits in different cavities. In such an architecture, quantum state
engineering and manipulation may involve not only qubits in the same cavity
\textit{but also qubits distributed in different cavities} [4,5]. The
ability to prepare quantum entangled states of qubits located in different
cavities and to perform nonlocal quantum operations on qubits in different
cavities is a prerequisite to realize large-scale QIP based on cavity or
circuit QED [6,7].

Greenberger-Horne-Zeilinger (GHZ) entangled states play a key role in
quantum communication and QIP. To give just a few examples, QIP [8], quantum
communication [9-11], error-correction protocols [12,13], quantum metrology
[14], and high-precision spectroscopy [15,16] require entangling quantum
systems in a GHZ state. New systems and methods for preparing and measuring
GHZ states have therefore been sought intensively for a long time, and
remains a very active field of research. To date, GHZ states of 10 or more
qubits have been experimentally demonstrated in various systems. For
examples, experiments have reported the generation of GHZ states with 14
ionic qubits [17], 20 atomic qubits [18], 12 photonic qubits via a linear
optical setup [19], 18 qubits with six photons' three degrees of freedom
[20], and 10 superconducting (SC) qubits coupled to a single microwave
resonator [21]. Moreover, GHZ states of 18 SC qubits coupled to a single
cavity or resonator has recently been produced in experiments [22]
(hereafter, the terms cavity and resonator are used interchangeably).
Theoretically, based on cavity or circuit QED, a large number of theoretical
methods have been presented for creating multi-qubit GHZ states with various
quantum systems (e.g., atoms, quantum dots, SC qutrits, NV centers, etc.),
which are placed in a single cavity or coupled to a single resonator
[23-31]. Moreover, proposals have been presented to entangle qubits
distributed in different cavities [32-42]. Note that the previous methods
presented for entangling qubits in a single cavity or resonator may not be
applied to entangle qubits that are distributed in different cavities, and
the previous proposals for entangling qubits in different cavities are not
universal, which depend on the specific cavity-system architecture and the
way in which the cavities are connected.

Motivated by the above, we present an efficient method to prepare GHZ states
of $N$-group qubits distributed in a $N$-cavity system. The multi-qubit GHZ
states are generated by using qutrits (three-level quantum systems) placed
in cavities or embedded in resonators. Here, the two logic states of a qubit
are represented by the two lowest levels of a qutrit placed in a cavity,
while the third higher energy level of each qutrit is utilized to facilitate
the coherent manipulation. By using this proposal, we show that given the
initial GHZ state of the cavities is prepared, the $N$-group qubits can be
deterministically prepared in a GHZ state with a 3-step operation only.
The procedure for creating the GHZ state of qubits works for a 1D
(one-dimensional), 2D, or 3D cavity-based quantum network (Fig.~1).
Moreover, it does not depend on in which way the cavities are connected
(e.g., via optical fibers or other auxiliary systems). This proposal is
quite general and can be used to create GHZ states of multiple groups of
qubits, by using natural atoms or artificial atoms (e.g., quantum dots, NV
centers, various SC qutrits, etc.) distributed in different cavities.\

Other advantages of this proposal are: (i) The GHZ state is prepared in a
deterministic way because neither measurement on the state of qutrits nor
measurement on the state of the cavities is needed; (ii) The GHZ-state
preparation time is independent of the number of qubits and thus does not
increase with the number of qubits; and (iii) The third level $\left\vert
f\right\rangle $ of the qutrits is not occupied during the entire operation,
thus decoherence from the higher energy levels of the qutrits is greatly
suppressed.

As an example, we further discuss the experimental feasibility of the proposal, based on circuit QED.
Our numerical simulations show that within current circuit QED technology, it is
feasible to produce GHZ states of four groups of SC transmon qubits, each group containing
three transmon qubits and the four groups distributed in four one-dimensional transmission line resonators (TLRs) arranged
in an array. By increasing the number of resonators, GHZ states of more groups of SC qubits can be
created experimentally.

This paper is organized as follows. Sec. II introduces basic theory. Sec.
III shows how to generate GHZ states of $N$-group qubits distributed in $N$%
cavities. Sec. IV investigates the experimental feasibility of preparing
GHZ states of four-group SC transmon qubits distributed in four TLRs arranged in an array. A
concluding summary is given in Sec. V.

\begin{center}
\textbf{II. BASIC THEORY }
\end{center}

\begin{figure}[tbp]
\begin{center}
\includegraphics[bb=0 459 553 800, width=12.0 cm, clip]{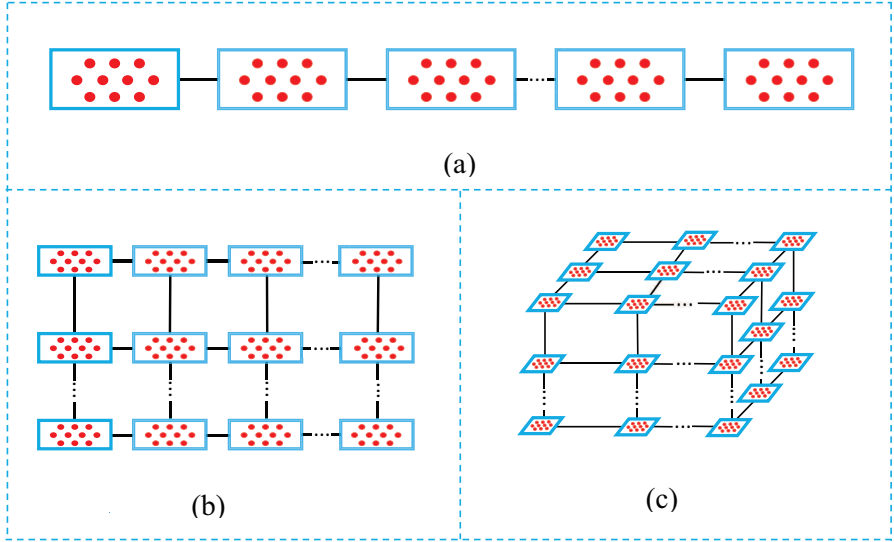} \vspace*{%
-0.08in}
\end{center}
\caption{(color online) (a) 1D cavity-based quantum network. (b) 2D
cavity-based quantum network. (c) 3D cavity-based quantum network. In
(a,b,c), each short line represents an optical fiber or other auxiliary
system, which is used to couple two adjacent cavities. In addition, each
cavity is a 1D or 3D cavity, hosting one group of qutrits (red dots).}
\label{fig:1}
\end{figure}

\begin{figure}[tbp]
\begin{center}
\includegraphics[bb=101 282 520 513, width=10.5 cm, clip]{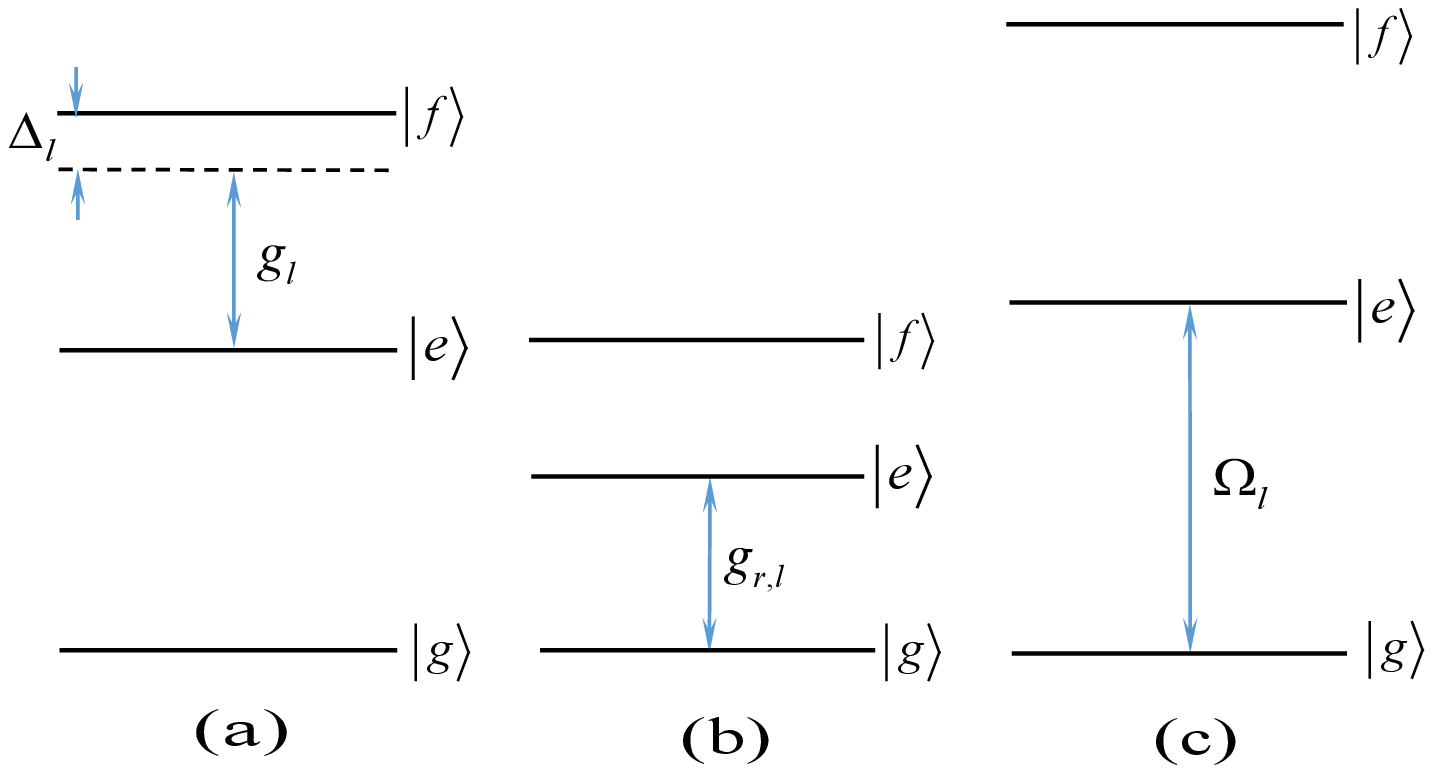} \vspace*{%
-0.08in}
\end{center}
\caption{(color online) (a) Illustration of the dispersive interaction
between cavity $l$ and the $\left\vert e\right\rangle $ $\leftrightarrow $ $\left\vert
f\right\rangle $ transition of qutrits $\left\{ 1_{l},2_{l},...,\left(
m-1\right) _{l}\right\} $, with coupling constant $g_{l} $
and detuning $\Delta _{l}=\protect\omega _{fe}-\protect\omega _{c_{l}}>0$.
Here, $\protect\omega_{fe}$ is the $\left\vert e\right\rangle $ $%
\leftrightarrow $ $\left\vert f\right\rangle $ transition frequency of the
qutrits and $\protect\omega _{c_{l}}$ is the frequency of cavity $l$. (b)
Illustration of the resonant interaction between cavity $l$ and the $\left\vert
g\right\rangle $ $\leftrightarrow $ $\left\vert e\right\rangle $ transition
of qutrit $m_{l}$ with coupling constant $g_{r,l}$. (c) Illustration of the resonant interaction between
a classical pulse and the $\left\vert g\right\rangle $ $%
\leftrightarrow $ $\left\vert e\right\rangle $ transition of qutrits $%
\left\{ 1_{l},2_{l},...,\left( m-1\right) _{l}\right\} $ in cavity $l$. Note
that the level structures in (a), (b), and (c) are different. The level
spacings of qutrits in (a) are adjusted such that $\left\vert e\right\rangle
\leftrightarrow \left\vert f\right\rangle $ transition is dispersively
coupled to cavity $l$. The level spacings in (b) are adjusted such that the $%
\left\vert g\right\rangle \leftrightarrow \left\vert e\right\rangle $
transition is resonant with cavity $l$. The level spacings in (c) are adjusted such that
qutrits are decoupled from cavity $l$ during the pulse. A blue double-arrow vertical
line in (a) and (b) represents the frequency of cavity $l$, while a blue
double-arrow vertical line in (c) represents the pulse frequency.}
\label{fig:2}
\end{figure}

Consider $N$ cavities ($1,2,...,N$) each hosting a group of qutrits
(Fig.~1). For simplicity, assume that each group contains $m$ qutrits. The $%
m $ qutrits hosted in cavity $l$ ($l=1,2,...,N$) are labelled as $1_{l},$ $%
2_{l},...,$ and $m_{l}$. The three levels of each qutrit are denoted as $%
\left\vert g\right\rangle ,$ $\left\vert e\right\rangle $ and $\left\vert
f\right\rangle $ (Fig. 2). As shown in the next section, the GHZ state
preparation requires: (i) Cavity $l$ dispersively interacting with the $%
\left\vert e\right\rangle \leftrightarrow $ $\left\vert f\right\rangle $
transition of each of qutrits $\left\{ 1_{l},2_{l},...,\left( m-1\right)
_{l}\right\} $ in cavity $l,$ (ii) Cavity $l$ resonantly interacting with
the $\left\vert g\right\rangle \leftrightarrow $ $\left\vert e\right\rangle $
transition of qutrit $m_{l}$ in cavity $l$, and (iii) A classical pulse
resonantly interacting with the $\left\vert g\right\rangle \leftrightarrow $
$\left\vert e\right\rangle $ transition of each of qutrits $\left\{
1_{l},2_{l},...,\left( m-1\right) _{l}\right\} $ in cavity $l$ ($l=1,2,...,N$%
). In the following, we will give a brief introduction to the state
evolution under these types of interaction.

\begin{center}
\textbf{A.} \textbf{Qutrit-cavity dispersive interaction}
\end{center}

Suppose that cavity $l$ is dispersively coupled to the $\left\vert
e\right\rangle $ $\leftrightarrow $ $\left\vert f\right\rangle $ transition
of each of qutrits $\left\{ 1_{l},2_{l},...,\left( m-1\right) _{l}\right\} $
with coupling strength $g_{l}$ and detuning $\Delta _{l}=\omega _{fe}-\omega
_{c_{l}}>0$, while highly detuned (decoupled) from other energy level
transitions [Fig.~2(a)]$.$ Here, $\omega _{fe}$ and $\omega _{c_{l}}$ are
the $\left\vert e\right\rangle $ $\leftrightarrow $ $\left\vert
f\right\rangle $ transition frequency of each\ qutrit and the frequency of
cavity $l,$ respectively. This condition can be met by prior adjustment of
the qutrit's level spacings or the frequency of cavity $l$. For instance,
the level spacings of superconducting qutrits can be rapidly (within $1\sim
3 $ ns) tuned [43,44]; the level spacings of NV centers can be readily
adjusted by changing the external magnetic field applied along the
crystalline axis of each NV center [45,46]; and the level spacings of
atoms/quantum dots can be adjusted by changing the voltage on the electrodes
around each atom/quantum dot [47]. In addition, the frequency for an optical
cavity can be changed in experiments [48], and the frequency of a microwave
cavity can be rapidly adjusted with a few nanoseconds [49,50].

Under the above assumptions, the Hamiltonian of the whole system in the
interaction picture and after the rotating wave approximation (RWA) is given by
(assuming $\hbar =1$)
\begin{equation}
H_{1}=\sum\limits_{l=1}^{N}g_{l}e^{i\Delta _{l}t}\hat{a}_{l}S_{fe,l}^{+}+%
\text{H.c.,}
\end{equation}%
where $S_{fe,l}^{+}=\sum\limits_{j=1}^{m-1}\left\vert f\right\rangle
_{j_{l}}\left\langle e\right\vert $, and $\hat{a}_{l}$ is the photon
annihilation operator of the cavity $l$ ($l=1,2,...,N$). In Eq. (1), we
assume that the coupling strength $g_{l}$ between cavity $l$ and the $%
\left\vert e\right\rangle $ $\leftrightarrow $ $\left\vert f\right\rangle $
transition is the same for all of qutrits $\left\{ 1_{l},2_{l},...,\left(
m-1\right) _{l}\right\} .$

Under the large detuning condition $\Delta _{l}\gg g_{l}\
(l=1,2,...,N),$ we can obtain the following effective Hamiltonian [51--53]
\begin{equation}
H_{\mathrm{eff}}=\sum\limits_{l=1}^{N}\lambda _{l}\left( S_{f,l}\hat{a}_{l}%
\hat{a}_{l}^{+}-S_{e,l}\hat{a}_{l}^{+}\hat{a}_{l}+\sum_{j,k=1;j\neq
k}^{m-1}\left\vert f\right\rangle _{j_{l}}\left\langle e\right\vert \otimes
\left\vert e\right\rangle _{k_{l}}\left\langle f\right\vert \right)
\end{equation}%
where $S_{f,l}=\sum\limits_{j=1}^{m-1}\left\vert f\right\rangle
_{j_{l}}\left\langle f\right\vert ,$ $S_{e,l}=\sum\limits_{j=1}^{m-1}\left%
\vert e\right\rangle _{j_{l}}\left\langle e\right\vert ,$ and $\lambda
_{l}=g_{l}^{2}/\Delta _{l}.$ Here, the first (second) term is an ac-Stark
shift of the level $\left\vert f\right\rangle $\ ($\left\vert e\right\rangle
$) induced by cavity $l$. The last term represents the \textquotedblleft
dipole\textquotedblright\ coupling between the $j$th and the $k$th qutrits
in cavity $l$, mediated by cavity $l$. When the level $\left\vert
f\right\rangle $ of each qutrit is not occupied, the Hamiltonian (2) reduces
to
\begin{equation}
H_{\mathrm{eff}}=-\sum\limits_{l=1}^{N}\lambda _{l}S_{e,l}\hat{a}_{l}^{+}%
\hat{a}_{l}.
\end{equation}%
Under this Hamiltonian, one can easily find that the following state evolution%
\begin{equation}
\begin{array}{c}
\left\vert g\right\rangle _{j_{l}}\left\vert 0\right\rangle _{c_{l}} \\
\left\vert e\right\rangle _{j_{l}}\left\vert 0\right\rangle _{c_{l}} \\
\left\vert g\right\rangle _{j_{l}}\left\vert 1\right\rangle _{c_{l}} \\
\left\vert e\right\rangle _{j_{l}}\left\vert 1\right\rangle _{c_{l}}%
\end{array}%
\rightarrow
\begin{array}{c}
\left\vert g\right\rangle _{j_{l}}\left\vert 0\right\rangle _{c_{l}} \\
\left\vert e\right\rangle _{j_{l}}\left\vert 0\right\rangle _{c_{l}} \\
\left\vert g\right\rangle _{j_{l}}\left\vert 1\right\rangle _{c_{l}} \\
e^{i\lambda _{l}t}\left\vert e\right\rangle _{j_{l}}\left\vert
1\right\rangle _{c_{l}}%
\end{array}%
.
\end{equation}%
applies to each of qutrits $\left\{ 1_{l},2_{l},...,\left( m-1\right)
_{l}\right\} $ in cavity $l$ simultaneously ($l=1,2,...,N$). Note that the subscript $j_l$ involved in
Eq. (4) is $1_l,2_l,...,$or $(m-1)_l$ $(l=1,2,...,N)$.

\begin{center}
\textbf{B. Qutrit-cavity resonant interaction}
\end{center}

Consider that cavity $l$ is resonant with the $\left\vert g\right\rangle $ $%
\leftrightarrow $ $\left\vert e\right\rangle $ transition of qutrit $m_{l}$ $%
(l=1,2,...,N)$ [Fig. 2(b)]. The Hamiltonian in the interaction picture and
after the RWA is given by
\begin{equation}
H_{2}=g_{r,l}\hat{a}_{l}\left\vert e\right\rangle _{m_{l}}\left\langle
g\right\vert +\text{H.c.},
\end{equation}%
where $g_{r,l}$ is the resonant coupling constant of cavity $l$ with the $%
\left\vert g\right\rangle $ $\leftrightarrow $ $\left\vert e\right\rangle $
transition of qutrit $m_{l}.$ Under this Hamiltonian, we can obtain the
state evolution
\begin{equation}
\left\vert g\right\rangle _{m_{l}}\left\vert 1\right\rangle
_{c_{l}}\rightarrow \cos g_{r,l}t\left\vert g\right\rangle
_{m_{l}}\left\vert 1\right\rangle _{c_{l}}-i\sin g_{r,l}t\left\vert
e\right\rangle _{m_{l}}\left\vert 0\right\rangle _{c_{l}},
\end{equation}%
while the state $\left\vert g\right\rangle _{m_{l}}\left\vert 0\right\rangle
_{c_{l}}$ remains unchanged.

\begin{center}
\textbf{C.} \textbf{Qutrit-pulse resonant interaction}
\end{center}

Assume that a classical pulse is resonant with the $\left\vert
g\right\rangle $ $\leftrightarrow $ $\left\vert e\right\rangle $ transition
of each of qutrits $\left\{ 1_{l},2_{l},...,\left( m-1\right) _{l}\right\} $
in cavity $l$ [Fig. 2(c)]. The Hamiltonian in the interaction picture and
after making the RWA is given by
\begin{equation}
H_{3}=\Omega _{l}e^{-i\phi }S_{eg,l}^{+}+\text{H.c.},
\end{equation}%
where $S_{eg,l}^{+}=\sum\limits_{j=1}^{m-1}\left\vert e\right\rangle
_{j_{l}}\left\langle g\right\vert ,$ $\phi $ is the pulse initial phase and $%
\Omega _{l}$ is the pulse Rabi frequency. Under this Hamiltonian, we can
easily obtain the following state rotation
\begin{eqnarray}
\left\vert g\right\rangle _{j_{l}} &\rightarrow &\cos \Omega _{l}t\left\vert
0\right\rangle -ie^{-i\phi }\sin \Omega _{l}t\left\vert 1\right\rangle ,
\notag \\
\left\vert e\right\rangle _{j_{l}} &\rightarrow &-ie^{i\phi }\sin \Omega
_{l}t\left\vert 0\right\rangle +\cos \Omega _{l}t\left\vert 1\right\rangle ,
\end{eqnarray}%
for qutrit $j_{l}$ ($j=1,2,...,m-1$).

The results (4), (6) and (8) will be applied for the GHZ state preparation,
as shown in the next section.

\begin{center}
\textbf{III. PREPARATION OF GHZ STATES OF }$\mathbf{N}$\textbf{-GROUP QUBITS
IN }$\mathbf{N}$\textbf{\ CAVITIES}
\end{center}

Assume that the $N$ cavities are initially prepared in a GHZ state $\alpha
\left\vert 0\right\rangle _{c_{1}}\left\vert 0\right\rangle
_{c_{2}}...\left\vert 0\right\rangle _{c_{N}}+\beta \left\vert
1\right\rangle _{c_{1}}\left\vert 1\right\rangle _{c_{2}}...\left\vert
1\right\rangle _{c_{N}}$ ($\left\vert \alpha \right\vert ^{2}+\left\vert
\beta \right\vert ^{2}=1,\alpha \neq 0,$ $\beta \neq 0$). In addition,
assume that qutrit $m_{l}$ in cavity $l$ is in the state $\left\vert
g\right\rangle $ while each of the remaining qutrits $\left\{
1_{l},2_{l},...,\left( m-1\right) _{l}\right\} $ in cavity $l$ is in the
state $\frac{1}{\sqrt{2}}$ $\left( \left\vert g\right\rangle +\left\vert
e\right\rangle \right) $, which can be prepared by applying a classical $\pi
$ pulse resonant with the $\left\vert g\right\rangle $ $\leftrightarrow $ $%
\left\vert e\right\rangle $ transition of the qutrits each initially in the
state $\left\vert g\right\rangle .$ Hereafter, define $\left\vert \pm
\right\rangle =\frac{1}{\sqrt{2}}$ $\left( \left\vert g\right\rangle \pm
\left\vert e\right\rangle \right) .$ The initial state of the whole system
is thus given by%
\begin{eqnarray}
&&\left( \alpha \left\vert 0\right\rangle _{c_{1}}\left\vert 0\right\rangle
_{c_{2}}...\left\vert 0\right\rangle _{c_{N}}+\beta \left\vert
1\right\rangle _{c_{1}}\left\vert 1\right\rangle _{c_{2}}...\left\vert
1\right\rangle _{c_{N}}\right)  \notag \\
&&\otimes \prod\limits_{j=1}^{m-1}\left\vert +\right\rangle
_{j_{1}}\prod\limits_{j=1}^{m-1}\left\vert +\right\rangle
_{j_{2}}...\prod\limits_{j=1}^{m-1}\left\vert +\right\rangle _{j_{N}}\otimes
\left\vert g\right\rangle _{m_{1}}\left\vert g\right\rangle
_{m_{2}}...\left\vert g\right\rangle _{m_{N}},
\end{eqnarray}%
where the subscripts $j_{1},j_{2},...,j_{N}$ represent the $j$th qutrit in
cavity $1$, cavity $2$, ..., cavity $N$ respectively; and $%
m_{1},m_{2},...m_{N}$ represent the $m$-th qutrit (i.e., qutrit $m$) in
cavity $1$, cavity $2$, ..., cavity $N$ respectively.

\begin{figure}[tbp]
\begin{center}
\includegraphics[bb=109 284 640 458, width=12.5 cm, clip]{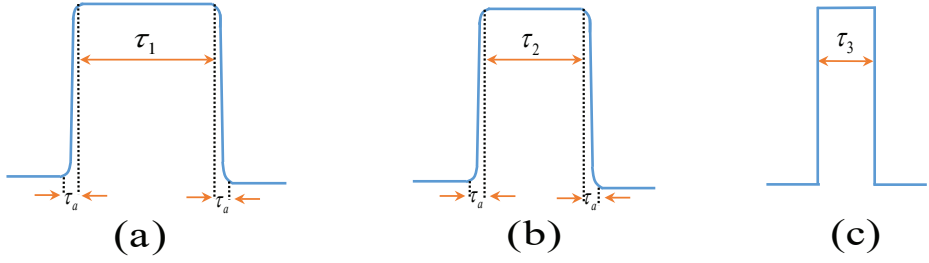} \vspace*{%
-0.08in}
\end{center}
\caption{(color online) (a) Sequence of operations for step 1. (b) Sequence
of operations for step 2. (c) Sequence of operations for step 3. Here, $%
\protect\tau _{1}$ and $\protect\tau _{2}$ are the qutrit-cavity interaction
times, while $\protect\tau _{3}$ is the qutrit-pulse interaction time, as
described in the text. In addition, $\protect\tau _{a}$ is the typical time
required to adjust the qutrit level spacings. Note that the operation sequence in
(a)-(c) follows from left to right.}
\label{fig:3}
\end{figure}

All qutrits are initially decoupled from their respective cavities. The
procedure for preparing the $N$-group qubits in a GHZ state is listed below:

Step 1. Keep qutrit $m_{l}$ decoupled from cavity $l$ but adjust the level
spacing of qutrits $\left\{ 1_{l},2_{l},...,\left( m-1\right) _{l}\right\} $
in cavity $l$ to obtain an effective Hamiltonian described by Eq.~(3).
According to Eq. (4), the state (9) evolves as follows
\begin{eqnarray}
&&\left[ \alpha \left\vert 0\right\rangle _{c_{1}}\left\vert 0\right\rangle
_{c_{2}}...\left\vert 0\right\rangle _{c_{N}}\otimes
\prod\limits_{j=1}^{m-1}\left\vert +\right\rangle
_{j_{1}}\prod\limits_{j=1}^{m-1}\left\vert +\right\rangle
_{j_{2}}...\prod\limits_{j=1}^{m-1}\left\vert +\right\rangle _{j_{N}}\right.
\notag \\
&&\left. +\beta \left\vert 1\right\rangle _{c_{1}}\left\vert 1\right\rangle
_{c_{2}}...\left\vert 1\right\rangle _{c_{N}}\prod\limits_{j=1}^{m-1}\frac{%
\left( \left\vert g\right\rangle _{j_{1}}+e^{i\lambda _{1}t}\left\vert
e\right\rangle _{j_{1}}\right) }{\sqrt{2}}\prod\limits_{j=1}^{m-1}\frac{%
\left( \left\vert g\right\rangle _{j_{2}}+e^{i\lambda _{2}t}\left\vert
e\right\rangle _{j_{2}}\right) }{\sqrt{2}}...\prod\limits_{j=1}^{m-1}\frac{%
\left( \left\vert g\right\rangle _{j_{N}}+e^{i\lambda _{N}t}\left\vert
e\right\rangle _{j_{N}}\right) }{\sqrt{2}}\right]  \notag \\
&&\otimes \left\vert g\right\rangle _{m_{1}}\left\vert g\right\rangle
_{m_{2}}...\left\vert g\right\rangle _{m_{N}}.
\end{eqnarray}%
By setting $\lambda _{1}=\lambda _{2}=...=\lambda _{N}=\lambda $ and for $%
t=\tau _{1}=\pi /\lambda ,$ the state (10) becomes
\begin{eqnarray}
&&\left( \alpha \left\vert 0\right\rangle _{c_{1}}\left\vert 0\right\rangle
_{c_{2}}...\left\vert 0\right\rangle _{c_{N}}\otimes
\prod\limits_{j=1}^{m-1}\left\vert +\right\rangle
_{j_{1}}\prod\limits_{j=1}^{m-1}\left\vert +\right\rangle
_{j_{2}}...\prod\limits_{j=1}^{m-1}\left\vert +\right\rangle _{j_{N}}\right.
\notag \\
&&\left. +\beta \left\vert 1\right\rangle _{c_{1}}\left\vert 1\right\rangle
_{c_{2}}...\left\vert 1\right\rangle
_{c_{N}}\prod\limits_{j=1}^{m-1}\left\vert -\right\rangle
_{j_{1}}\prod\limits_{j=1}^{m-1}\left\vert -\right\rangle
_{j_{2}}...\prod\limits_{j=1}^{m-1}\left\vert -\right\rangle _{j_{N}}\right)
\notag \\
&&\otimes \left\vert g\right\rangle _{m_{1}}\left\vert g\right\rangle
_{m_{2}}...\left\vert g\right\rangle _{m_{N}}.
\end{eqnarray}
Then, adjust the level spacings of qutrits $\left\{ 1_{l},2_{l},...,\left( m-1\right) _{l}\right\} $
such that they are decoupled from cavity $l$. The operation sequence for this step of operation is illustrated
in Fig.~3(a).

Step 2. Adjust the level spacing of qutrit $m_{l}$ in cavity $l$ such that
the $\left\vert g\right\rangle $ $\leftrightarrow $ $\left\vert
e\right\rangle $ transition of qutrit $m_{l}$ is resonant with cavity $l$
(with a resonant coupling constant $g_{r,l}$). After an interaction time $\tau _{2}=\pi /\left(
2g_{r,l}\right) $, we have $\left\vert 1\right\rangle
_{c_{l}}\left\vert g\right\rangle _{m_{l}}\rightarrow -i\left\vert
0\right\rangle _{c_{l}}\left\vert e\right\rangle _{m_{l}}$ according to Eq.~(6). Thus, the state
(11) becomes
\begin{eqnarray}
&&\left( \alpha \prod\limits_{j=1}^{m-1}\left\vert +\right\rangle
_{j_{1}}\prod\limits_{j=1}^{m-1}\left\vert +\right\rangle
_{j_{2}}...\prod\limits_{j=1}^{m-1}\left\vert +\right\rangle _{j_{N}}\otimes
\left\vert g\right\rangle _{m_{1}}\left\vert g\right\rangle
_{m_{2}}...\left\vert g\right\rangle _{m_{N}}\right.  \notag \\
&&\left. +\left( -i\right) ^{N}\beta \prod\limits_{j=1}^{m-1}\left\vert
-\right\rangle _{j_{1}}\prod\limits_{j=1}^{m-1}\left\vert -\right\rangle
_{j_{2}}...\prod\limits_{j=1}^{m-1}\left\vert -\right\rangle _{j_{N}}\otimes
\left\vert e\right\rangle _{m_{1}}\left\vert e\right\rangle
_{m_{2}}...\left\vert e\right\rangle _{m_{N}}\right)  \notag \\
&&\otimes \left\vert 0\right\rangle _{c_{1}}\left\vert 0\right\rangle
_{c_{2}}...\left\vert 0\right\rangle _{c_{N}}.
\end{eqnarray}%
To maintain the state (12), one should adjust the level spacing of qutrit $m_l$
such that it is decoupled from cavity $l.$ The
operation sequence for this step of operation is illustrated in Fig. 3(b).

Step 3. Apply a classical $\pi $ pulse (with an initial phase $\pi /2$) to
qutrit $j_{l}$ ($j=1,2,...,m-1$). The pulse is resonant with the $\left\vert
g\right\rangle $ $\leftrightarrow $ $\left\vert e\right\rangle $ transition
of qutrit $j_{l}$ for a duration time $\tau _{3}=\pi /\left( 2\Omega
_{l}\right) ,$ resulting in $\left\vert +\right\rangle _{j_{l}}\rightarrow
\left\vert g\right\rangle _{j_{l}}$ and $\left\vert -\right\rangle
_{j_{l}}\rightarrow -\left\vert e\right\rangle _{j_{l}}$ according to Eq.
(8). The state (12) thus becomes
\begin{equation}
\alpha \prod\limits_{j=1}^{m}\left\vert g\right\rangle
_{j_{1}}\prod\limits_{j=1}^{m}\left\vert g\right\rangle
_{j_{2}}...\prod\limits_{j=1}^{m}\left\vert g\right\rangle _{j_{N}}+e^{i\phi
}\beta \prod\limits_{j=1}^{m}\left\vert e\right\rangle
_{j_{1}}\prod\limits_{j=1}^{m}\left\vert e\right\rangle
_{j_{2}}...\prod\limits_{j=1}^{m}\left\vert e\right\rangle _{j_{N}},
\end{equation}%
where $\phi =\left( m-3/2\right) N\pi .$ This state is a GHZ entangled state
for the $N$-group qubits in the $N$ cavities, with the two logic states of a
qubit being represented by the two lowest levels $\left\vert g\right\rangle $
and $\left\vert e\right\rangle $ of a qutrit. For $\left\vert \alpha
\right\vert =\left\vert \beta \right\vert =1/\sqrt{2},$ the state (13) is a
standard GHZ state with maximal entanglement. The operation sequence for
this step of operation is illustrated in Fig. 3(c).

In above, we have set $\lambda _{1}=\lambda _{2}=...=\lambda _{N}$, which
turns out into
\begin{equation}
\frac{g_{1}^{2}}{\Delta _{1}}=\frac{g_{2}^{2}}{\Delta _{2}}=...=\frac{%
g_{N}^{2}}{\Delta _{N}}.
\end{equation}%
This condition (14) can be readily met by adjusting the qutrits' positions in
the cavities, the qutrits' level spacings [43-47] or the cavity frequencies
[48-50].

From the above description, one can see:

(i) Because the same detuning $\Delta _{l}$ is set for each of qutrits $%
1_{l},2_{l},...,\left( m-1\right) _{l}$ in cavity $l$ ($l=1,2,...,N$), the
level spacings for qutrits $1_{l},2_{l},...,\left( m-1\right) _{l}$ can be
synchronously adjusted, e.g., via changing the common external parameters.

(ii) During the entire operation, the level $\left\vert f\right\rangle $
for all qutrits in each cavity is not occupied. Thus, decoherence due to
energy relaxation and dephasing of this higher energy level is greatly
suppressed.

(iii) Assume that both $g_{r,1},g_{r,2},...,g_{r,N}$ and $\Omega _{1},\Omega
_{2},...,\Omega _{N}$ are non-identical for different cavities. Thus, the
total operation time is
\begin{equation}
t_{op}=\pi /\lambda +\max \{\frac{\pi }{2g_{r,1}},\frac{\pi }{2g_{r,2}},...,%
\frac{\pi }{2g_{r,N}}\}+\max \{\frac{\pi }{2\Omega _{1}},\frac{\pi }{2\Omega
_{2}},...,\frac{\pi }{2\Omega _{N}}\}+4\tau _{d},
\end{equation}%
which is independent of the number of qubits and thus does not increase with
the number of qubits. Note that $\tau _{d}$ is the typical time required for
adjusting the level spacings of qutrits.

(iv) This proposal does not require measurement on the state of the qutrits
or the cavities. Thus, the GHZ state is created deterministically.

(v) The above operations have nothing to do with the manner in which the
cavities are connected. In this sense, the method presented here can be
applied to create GHZ states of the qubits distributed in a 1D, 2D, or 3D
cavity-based quantum network (Fig.~1), where the cavities can be connected
with optical fibers or other auxiliary systems.

(vi) When the $N$ cavities are initially prepared in another type of symmetrical GHZ
state $\alpha \left\vert 0\right\rangle _{c_{1}}\left\vert 0\right\rangle
_{c_{2}}...\left\vert 0\right\rangle _{c_{s}}\left\vert 1\right\rangle
_{c_{s+1}}\left\vert 1\right\rangle _{c_{s+2}}...\left\vert 1\right\rangle
_{c_{N}}+\beta \left\vert 1\right\rangle _{c_{1}}\left\vert 1\right\rangle
_{c_{2}}...\left\vert 1\right\rangle _{c_{s}}\left\vert 0\right\rangle
_{c_{s+1}}\left\vert 0\right\rangle _{c_{s+2}}...\left\vert 0\right\rangle
_{c_{N}},$ it is straightforward to show that by following the procedure
described above, the $N$-group qubits distributed in $N$ cavities will be
prepared in the following GHZ state
\begin{eqnarray}
&&\alpha \prod\limits_{j=1}^{m}\left\vert g\right\rangle
_{j_{1}}\prod\limits_{j=1}^{m}\left\vert g\right\rangle
_{j_{2}}...\prod\limits_{j=1}^{m}\left\vert g\right\rangle
_{j_{s}}\prod\limits_{j=1}^{m}\left\vert e\right\rangle
_{j_{s+1}}\prod\limits_{j=1}^{m}\left\vert e\right\rangle
_{j_{s+2}}...\prod\limits_{j=1}^{m}\left\vert e\right\rangle _{j_{N}}  \notag
\\
&&+\beta \prod\limits_{j=1}^{m}\left\vert e\right\rangle
_{j_{1}}\prod\limits_{j=1}^{m}\left\vert e\right\rangle
_{j_{2}}...\prod\limits_{j=1}^{m}\left\vert e\right\rangle
_{j_{s}}\prod\limits_{j=1}^{m}\left\vert g\right\rangle
_{j_{s+1}}\prod\limits_{j=1}^{m}\left\vert g\right\rangle
_{j_{s+2}}...\prod\limits_{j=1}^{m}\left\vert g\right\rangle _{j_{N}}.
\end{eqnarray}

(vii) The procedure described above can also be applied to create GHZ state
of $N-$ group qubits distributed in $N$ cavities in the case when the number
of qutrits in each group is different.

As a matter of fact, the condition (14) is unnecessary. For the case of $%
\lambda _{1}\neq \lambda _{2}\neq ...\neq \lambda _{N}$, the state (11)
resulting from the operation of step 1 described above cannot be achieved
by turning on/off the effective couplings of the qutrits with the $N$
cavities simultaneously. However, this state (11) can be obtained by
modifying the operation of step 1 as follows. First,
switch on the effective dispersive interaction of the qutrits $%
\{1_{l},2_{l},...,\left( m-1\right) _{l}\}$ with cavity $l$ at a proper time
$\tau _{l}$ = $t_{\max }-t_{l},$ by tuning the frequency of the qutrits $%
\{1_{l},2_{l},...,\left( m-1\right) _{l}\}$ or the frequency of cavity $l$
to have the proper $\Delta _{l},$ where $t_{\max }$ $=\max \{\pi /\left(
2\lambda _{1}\right) ,\pi /\left( 2\lambda _{2}\right) ,...,\pi /\left(
2\lambda _{N}\right) \}$ and $t_{l}=$ $\pi /\left( 2\lambda _{l}\right) $.
Then, switch off all the effective interactions of the qutrits with the $N$
cavities at the time $t_{\max },$ by tuning the frequency of the qutrits or
the frequency of the $N$ cavities such that the qutrits are decoupled from
the $N$ cavities.

In the above discussion, we have assumed that the coupling strength $g_{l}$
is identical for all of qutrits $\{1_{l},2_{l},...,\left( m-1\right) _{l}\}$
in cavity $l$ ($l=1,2,...,N)$. For the case of $g_{l}$\ varying with
different qutrits in cavity $l,$ this proposal is still valid as long as the
large detuning condition holds for individual qutrits, but the procedure may
become more complex because one will need to adjust the frequencies of
individual qutrits separately. Therefore, to simplify the experiments, it is
strongly suggested to design the sample with identical qutrit-cavity
coupling strength for qutrits in the same cavity.

To prepare the cavities in the GHZ state, two key ingredients are required.
One is the coupling between neighbor cavities. For optical cavities, this
can be obtained by using optical fibers to connect the neighbor cavities. In
addition, for microwave cavities or resonators, this can be achieved by
using solid-state auxiliary systems (e.g., superconducting qubits/qutrits,
quantum dots, or NV centers) to connect the neighbor cavities. The other is
decoupling of the intra-cavity atoms with the cavities. This can be realized
by adjusting the level spacings of the atoms or the frequencies of the
cavities such that the cavities are highly detuned (decoupled) from the
transitions between any two levels of the atoms. As discussed previously,
both level spacings of natural or artificial atoms and cavity frequencies
can be adjusted in experiments [43-50].

\begin{figure}[tbp]
\begin{center}
\includegraphics[bb=74 424 533 496, width=11.5 cm, clip]{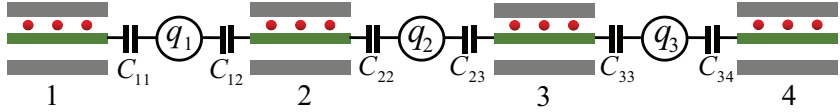} \vspace*{%
-0.08in}
\end{center}
\caption{(color online) 1D quantum network consisting of four
one-dimensional transmission line resonators (TLRs) arranged in an array.
Each TLR hosts three SC transmon qutrits (red dots), and adjacent TLRs are
coupled through SC transmon qutrits ($q_1,q_2,q_3$).}
\label{fig:4}
\end{figure}

\begin{center}
\textbf{IV. POSSIBLE EXPERIMENTAL IMPLEMENTATION}
\end{center}

In above, a general type of qubit is considered and a qubit is formed by the
two lowest levels of a qutrit. Circuit QED consists of microwave cavities
and superconducting (SC) qubits, which is an analogue of cavity QED and has
been considered as one of the leading candidates for QIP [54-60]. As an
example, let us consider a setup, which consists of four TLRs, each hosting
three SC transmon qutrits, connected through the coupler SC transmon qutrits
($q_{1},q_{2},q_{3}$), and arranged in an array (Fig. 4). The three SC transmon qutrits placed in
cavity $l$ are labelled as $1_l,2_l$, and $3_l$ ($l=1,2,3,4$). In the following,
we will give a discussion on the experimental feasibility of preparing a GHZ
state of the four-group SC transmon qubits distributed in the four TLRs (Fig. 4).

Let us first give some explanation on transmon qutrits and transmon qubits.
A transmon qutrit has a ladder-type three level structure as shown in Fig.
2, while a transmon qubit considered here is formed by the two lowest levels
$\left\vert g\right\rangle $ and $\left\vert e\right\rangle $ of a transmon
qutrit. In other words, when the third level $\left\vert f\right\rangle $ of
a transmon qutrit is dropped off (Fig. 2), the transmon qutrit reduces to a
transmon qubit. As is well known, a transom qubit is an artificial two-level
atom, whose Hamiltonian takes the same form as the Hamiltonian of a natural
two-level atom, i.e., $H=\omega _{0}\sigma _{z}$, where $\omega _{0}$ is the
transition frequency of the atom, and $\sigma _{z}=\left\vert e\right\rangle
\left\langle e\right\vert -\left\vert g\right\rangle \left\langle
g\right\vert $ is the Pauli operator. Based on the discussion here, one can
see that the three tranmon qutrits (red dots in Fig. 4) placed in a TLR
correspond to three transmon qubits (i.e., one group of qubits). Thus, the
four groups of transmon qutrits placed in the four TLRs correspond to the
four groups of SC transmon qubits. For convenience, in the following we will use
{\it the terms ``cavity" and ``resonator"} interchangeably.

\begin{figure}[tbp]
\begin{center}
\includegraphics[bb=99 284 520 536, width=10.5 cm, clip]{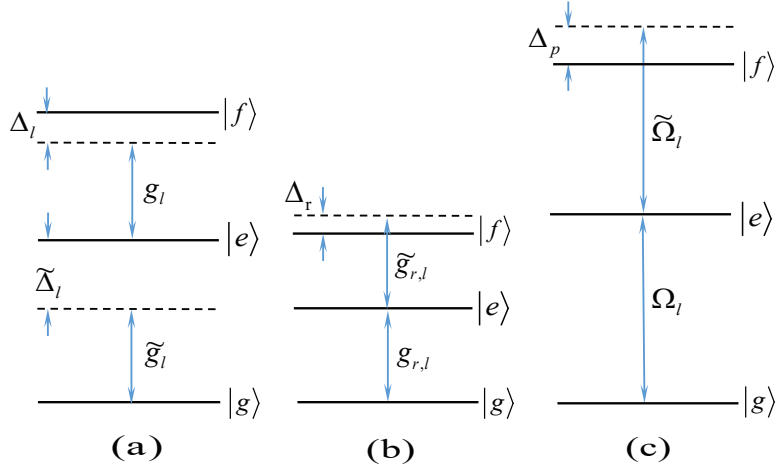} \vspace*{%
-0.08in}
\end{center}
\caption{(color online) (a) Dispersive interaction between cavity $l$ and the $%
\left\vert e\right\rangle $ $\leftrightarrow $ $\left\vert f\right\rangle $
transition of qutrits $\left\{ 1_{l},2_{l}\right\} $ with coupling strength $%
g_{l}$ and detuning $\Delta _{l}=\protect\omega _{fe}-\protect\omega %
_{c_{l}}>0$, as well as the unwanted off-resonant interaction between cavity $l$ and
the $\left\vert g\right\rangle $ $%
\leftrightarrow $ $\left\vert e\right\rangle $ transition of qutrits $%
\left\{ 1_{l},2_{l}\right\} $ with coupling strength $\widetilde{g}_{l}$ and
detuning $\widetilde{\Delta }_{l}=\protect\omega _{eg}-\protect\omega %
_{c_{l}}>0$. (b) Resonant interaction between cavity $l$ and the $\left\vert g\right\rangle
\leftrightarrow \left\vert e\right\rangle $ transition of qutrit $3_{l}$
with coupling constant $g_{r,l}$, as well as the unwanted off-resonant interaction between
cavity $l$ and the $\left\vert
e\right\rangle \leftrightarrow \left\vert f\right\rangle $ transition of
qutrit $3_{l}$ with coupling constant $\widetilde{g}_{r,l}$ and detuning $%
\Delta _{r,l}$. (c) Resonant interaction between a classical pulse and the $\left\vert
g\right\rangle \leftrightarrow \left\vert e\right\rangle $ transition of
qutrits $\left\{ 1_{l},2_{l}\right\} $ with Rabi frequency $\Omega _{l}$, as well as the unwanted
off-resonant interaction between the pulse and the $\left\vert e\right\rangle \leftrightarrow
\left\vert f\right\rangle $ transition of qutrits $\left\{
1_{l},2_{l}\right\} $ with Rabi frequency $\widetilde{\Omega }_{l}$ and
detuning $\Delta _{p}=\omega_{fe}-\omega_p$. Here, $\omega_p$ is the pulse frequency.}
\label{fig:5}
\end{figure}

From the description given in the previous section, one can see that three
basic interactions are used in the preparation of the GHZ states, i.e., the
three basic interactions described by the Hamiltonians $H_{1}, H_{2},$ and $%
H_{3}$ described above. With the unwanted interaction and the inter-cavity
crosstalk being considered, these Hamiltonians are modified as follows:

(i) $H_{1}^{\prime }=H_{1}+\delta \!H_{1}+\varepsilon $, where $\delta
\!H_{1}$ describes the unwanted interaction of cavity $l$ with the $%
\left\vert g\right\rangle \leftrightarrow \left\vert e\right\rangle $
transition of qutrits $\left\{ 1_{l},2_{l}\right\} $ in cavity $l$ ($l=1,2,3,4$) [Fig. 5(a)].
The expression of $\delta \!H_{1}$ is given by
\begin{equation}
\delta H_{1}=\sum\limits_{l=1}^{4}\widetilde{g}_{l}e^{i\widetilde{\Delta }%
_{l}t}\hat{a}_{l}S_{eg,l}^{+}+\text{H.c.,}
\end{equation}%
where $S_{eg,l}^{+}=\sum\limits_{j=1}^{2}\left\vert e\right\rangle
_{j_{l}}\left\langle g\right\vert $, $\widetilde{g}_{l}$ is the coupling
strength between cavity $l$ and the $\left\vert g\right\rangle
\leftrightarrow \left\vert e\right\rangle $ transition of qutrits $%
\left\{ 1_{l},2_{l}\right\} $, and $\widetilde{\Delta }_{l}=\omega _{eg}-\omega _{c_{l}}$ is the
detuning between the frequency of cavity $l$ and the $\left\vert
g\right\rangle \leftrightarrow \left\vert e\right\rangle $ transition
frequency of qutrits $\left\{ 1_{l},2_{l}\right\} $. In addition, $\varepsilon $
describes the inter-cavity crosstalk between the adjacent cavities, which is
given by%
\begin{equation}
\varepsilon =g_{12}e^{i\Delta _{12}t}\hat{a}_{1}^{+}\hat{a}%
_{2}+g_{23}e^{i\Delta _{23}t}\hat{a}_{2}^{+}\hat{a}_{3}+g_{34}e^{i\Delta
_{34}t}\hat{a}_{3}^{+}\hat{a}_{4}+\text{H.c.},
\end{equation}%
where $\Delta _{j(j+1)}=\omega _{c_{j}}-\omega _{c_{j+1}}=\Delta _{j+1}-\Delta _{j}$ $(j=1,2,3)$, $
g_{j(j+1)}$ is the crosstalk strength between the two neighbor cavities $j$ and $j+1$ $(j=1,2,3).$ Note
that when compared to the crosstalk between the adjacent cavities, the
crosstalk between non-adjacent cavities (i.e., cavities $1$ and $3$,
cavities $1$ and $4$, and cavities $2$ and $4$) are negligible.

(ii) $H_{2}^{\prime }=H_{2}+\delta \!H_{2}+\varepsilon ,$ where $\delta
\!H_{2}$ describes the unwanted interaction between cavity $l$ and the $%
\left\vert e\right\rangle \leftrightarrow \left\vert f\right\rangle $
transition of qutrit $3_{l}$ in cavity $l$ ($l=1,2,3,4$) [Fig. 5(b)]$.$ The
expression of $\delta \!H_{2}$ is given by
\begin{equation}
\delta \!H_{2}=\widetilde{g}_{r,l}e^{i\Delta _{r,l}t}\hat{a}_{l}\left\vert
f\right\rangle _{3_{l}}\left\langle e\right\vert +\text{H.c.}
\end{equation}%
where $\widetilde{g}_{r}$ is the off-resonant coupling strength between
cavity $l$ and the $\left\vert e\right\rangle \leftrightarrow \left\vert
f\right\rangle $ transition of qutrit $3_{l}$ in cavity $l,$ and $\Delta
_{r,l}=\omega _{fe}-\omega _{c_{l}}$ is the detuning between the frequency
of cavity $l$ and the $\left\vert e\right\rangle \leftrightarrow \left\vert
f\right\rangle $ transition frequency of qutrit $3_{l}.$

(iii) $\widetilde{H}_{3}=H_{3}+\delta \!H_{3}+\varepsilon ,$ where $\delta
\!H_{3}$ describes the unwanted interaction between the pulse and the $%
\left\vert e\right\rangle \leftrightarrow \left\vert f\right\rangle $
transition of $\left\{ 1_{l},2_{l}\right\} $ ($l=1,2,3,4$) [Fig. 5(c)]$%
. $ The expression of $\delta \!H_{3}$ is given by
\begin{equation}
\delta \!H_{3}=\widetilde{\Omega }_{l}e^{-i\phi }e^{-i\Delta
_{p}t}S_{fe,l}^{+}+\text{H.c.}
\end{equation}%
where $S_{fe,l}^{+}=\sum\limits_{j=1}^{2}\left\vert f\right\rangle
_{j_{l}}\left\langle e\right\vert ,$ $\widetilde{\Omega }_{l}$ is the pulse
Rabi frequency associated with the $\left\vert e\right\rangle
\leftrightarrow \left\vert f\right\rangle $ transition of the qutrits, and $%
\Delta _{p}=\omega _{fe}-\omega _{p}=\omega _{fe}-\omega _{eg}$ is the
detuning between the pulse frequency $\omega_p$ and the $\left\vert e\right\rangle
\leftrightarrow \left\vert f\right\rangle $ transition frequency of the
qutrits.

It should be mentioned that the $\left\vert g\right\rangle \leftrightarrow
\left\vert f\right\rangle $ transition induced by the pulse or the cavities
is negligible because $\omega _{eg},\omega _{fe}\ll \omega _{fg}$ (Fig.~2).
For simplicity, we also assume that the effect of the qutrit decoherence and
the cavity decay during the adjustment of the qutrit level spacings is
negligible because for transmon qutrits the level spacings can be rapidly
adjusted.

After taking into account the qutrit decoherence and the cavity decay, the
system dynamics, under the Markovian approximation, is determined by the
master equation
\begin{eqnarray}
\frac{d\rho }{dt} &=&-i\left[ H_{k}^{\prime },\rho \right]
+\sum\limits_{l=1}^{4}\kappa _{l}\mathcal{L}\left[ \hat{a}_{l}\right] +
\notag \\
&&+\gamma _{eg}\sum\limits_{l=1}^{4}\sum\limits_{j=1}^{3}\mathcal{L}\left[
\sigma _{eg,j_{l}}^{-}\right] +\gamma
_{fe}\sum\limits_{l=1}^{4}\sum\limits_{j=1}^{3}\mathcal{L}\left[ \sigma
_{fe,j_{l}}^{-}\right] +\gamma
_{fg}\sum\limits_{l=1}^{4}\sum\limits_{j=1}^{3}\mathcal{L}\left[ \sigma
_{fg,j_{l}}^{-}\right]  \notag \\
&&+\gamma _{\varphi ,e}\sum\limits_{l=1}^{4}\sum\limits_{j=1}^{3}\left(
\sigma _{ee,j_{l}}\rho \sigma _{ee,j_{l}}-\sigma _{ee,j_{l}}\rho /2-\rho
\sigma _{ee,j_{l}}/2\right)  \notag \\
&&+\gamma _{\varphi ,f}\sum\limits_{l=1}^{4}\sum\limits_{j=1}^{3}\left(
\sigma _{ff,j_{l}}\rho \sigma _{ff,j_{l}}-\sigma _{ff,j_{l}}\rho /2-\rho
\sigma _{ff,j_{l}}/2\right) ,
\end{eqnarray}%
where $H_{k}^{\prime }$ (with $k=1,2,3$) are the modified Hamiltonians $%
H_{1}^{\prime },$ $H_{2}^{\prime },$ and $H_{3}^{\prime }$ given above, $%
\mathcal{L}\left[ \Lambda \right] =\Lambda \rho \Lambda ^{+}-\Lambda
^{+}\Lambda \rho /2-\rho \Lambda ^{+}\Lambda /2$ (with $\Lambda =\hat{a}%
_{l},,\sigma _{fe,j_{l}}^{-},\sigma _{eg,j_{l}}^{-},\sigma _{fg,j_{l}}^{-}$%
), $\sigma _{fe,j_{l}}^{-}=\left\vert e\right\rangle _{j_{l}}\left\langle
f\right\vert ,$ $\sigma _{eg,j_{l}}^{-}=\left\vert g\right\rangle
_{j_{l}}\left\langle e\right\vert ,$ $\sigma _{fg,j_{l}}^{-}=\left\vert
g\right\rangle _{j_{l}}\left\langle f\right\vert ,$ $\sigma
_{ee,j_{l}}=\left\vert e\right\rangle _{j_{l}}\left\langle e\right\vert $,
and $\sigma _{ff,j_{l}}=\left\vert f\right\rangle _{j_{l}}\left\langle
f\right\vert .$ In addition, $\kappa _{l}$ is the decay rate of cavity $l$;\
$\gamma _{eg}$\ is the energy relaxation rate for the level $\left\vert
e\right\rangle $\ associated with the decay path $\left\vert e\right\rangle
\rightarrow \left\vert g\right\rangle $; $\gamma _{fe}$\ ($\gamma _{fg}$) is
the relaxation rate for the level $\left\vert f\right\rangle $ related to
the decay path $\left\vert f\right\rangle \rightarrow \left\vert
e\right\rangle $ ($\left\vert f\right\rangle \rightarrow \left\vert
g\right\rangle $); $\gamma _{\varphi ,e}$ ($\gamma _{\varphi ,f}$) is the
dephasing rate of the level $\left\vert e\right\rangle $ ($\left\vert
f\right\rangle $)\textbf{. }

The fidelity of the operation is given by $\mathcal{F}=\sqrt{\left\langle
\psi _{id}\right\vert \rho \left\vert \psi _{id}\right\rangle },$ where $%
\left\vert \psi _{id}\right\rangle $ is the ideal output state given by%
\begin{equation}
\frac{1}{\sqrt{2}}\left( \prod\limits_{j=1}^{3}\left\vert g\right\rangle
_{j_{1}}\prod\limits_{j=1}^{3}\left\vert g\right\rangle
_{j_{2}}\prod\limits_{j=1}^{3}\left\vert e\right\rangle
_{j_{3}}\prod\limits_{j=1}^{3}\left\vert e\right\rangle
_{j_{4}}+\prod\limits_{j=1}^{3}\left\vert e\right\rangle
_{j_{1}}\prod\limits_{j=1}^{3}\left\vert e\right\rangle
_{j_{2}}\prod\limits_{j=1}^{3}\left\vert g\right\rangle
_{j_{3}}\prod\limits_{j=1}^{m}\left\vert g\right\rangle _{j_{4}}\right)
\otimes \prod\limits_{l=1}^{4}\left\vert 0\right\rangle _{c_{l}},
\end{equation}%
when the four TLRs are initially in the GHZ state $\frac{1}{%
\sqrt{2}}\left( \left\vert 0\right\rangle _{c_{1}}\left\vert 0\right\rangle
_{c_{2}}\left\vert 1\right\rangle _{c_{3}}\left\vert 1\right\rangle
_{c_{4}}+\left\vert 1\right\rangle _{c_{1}}\left\vert 1\right\rangle
_{c_{2}}\left\vert 0\right\rangle _{c_{3}}\left\vert 0\right\rangle
_{c_{4}}\right) $ (see the appendix for the details of preparing the four
TLRs in this GHZ state), while $\rho $ is the final density matrix obtained
by numerically solving the master equation.

We now numerically calculate the fidelity. For a transmon qutrit, the level
spacing anharmonicity $100\sim 720$ MHz was reported in experiments [61]. As
an example, consider $\Delta _{r,l}/2\pi =\Delta _{p}/2\pi =-\left(
\widetilde{\Delta }_{l}-\Delta _{l}\right) /2\pi =-0.7$ GHz. By choosing $%
\Delta _{1}/2\pi =\Delta _{3}/2\pi =100$ MHz and $\Delta _{2}/2\pi =\Delta
_{4}/2\pi =80$ MHz, we have $\Delta _{12}/2\pi =-20$ MHz, $\Delta _{23}/2\pi
=20$ MHz, and $\Delta _{34}/2\pi =-20$ MHz. With the choice of $\Delta
_{1},\Delta _{2},\Delta _{3},\Delta _{4}$ here, one has $g_{2}=g_{4}=\sqrt{%
\frac{4}{5}}g_{1}$ and $g_{3}=g_{1}$ according to Eq.~(14). For transmon
qutrits [62], $\widetilde{g}_{l}=g_{l}/\sqrt{2},$ $\widetilde{g}_{r,l}=\sqrt{%
2}g_{r,l},$ $\widetilde{\Omega }_{l}=\sqrt{2}\Omega _{l}.$ For simplicity,
we assume $g_{r,l}=\widetilde{g}_{l}.$ In addition, we choose $%
g_{12},g_{23},g_{34}=0.01\max \{g_{1},g_{2},g_{3}\},$ which is achievable in
experiments by a prior design of the sample with appropriate capacitances $%
c_{11},c_{12},c_{22,}c_{23},c_{33},c_{34}$ [63]. Other parameters used in
the numerical simulation are: (i) $\gamma _{eg}^{-1}=60$ $\mu $s, $\gamma
_{fg}^{-1}=150$ $\mu $s [64], $\gamma _{fe}^{-1}=30$ $\mu $s, $\gamma _{\phi
,e}^{-1}=\gamma _{\phi ,f}^{-1}=20$ $\mu $s, (ii) $\Omega _{l}/2\pi =45$
MHz. Here, we consider a rather conservative case for decoherence time of
the transmon qutrit [65,66]. For simplicity, we assume $\kappa
_{l}=\kappa$ in our numerical simulation ($l=1,2,3,4$).

By numerically solving the master equation (21), we plot Fig. 6 for $\kappa
^{-1}=10$ $\mu $s, which shows the fidelity versus $g_{1}.$ From Fig. 6,
one can see that for $g_{1}/2\pi \sim $ $14.15$ MHz, a high fidelity $\sim
90\%$ can be obtained. For the value of $g_{1}$ here$,$ $g_{2}/2\pi
,g_{4}/2\pi \sim 12.65$ MHz; $g_{3}\sim 14.15$ MHz; $g_{r,1}/2\pi
,g_{r,3}/2\pi \sim 10$ MHz; and $g_{r,2}/2\pi ,g_{r,4}/2\pi =8.95$ MHz$,$
which are readily available in experiments because a coupling strength $%
g/2\pi $ $\sim 360$ MHz has been reported for a transmon qutrit coupled to a
TLR [67,68].

\begin{figure}[tbp]
\begin{center}
\includegraphics[bb=0 0 440 286, width=11.5 cm, clip]{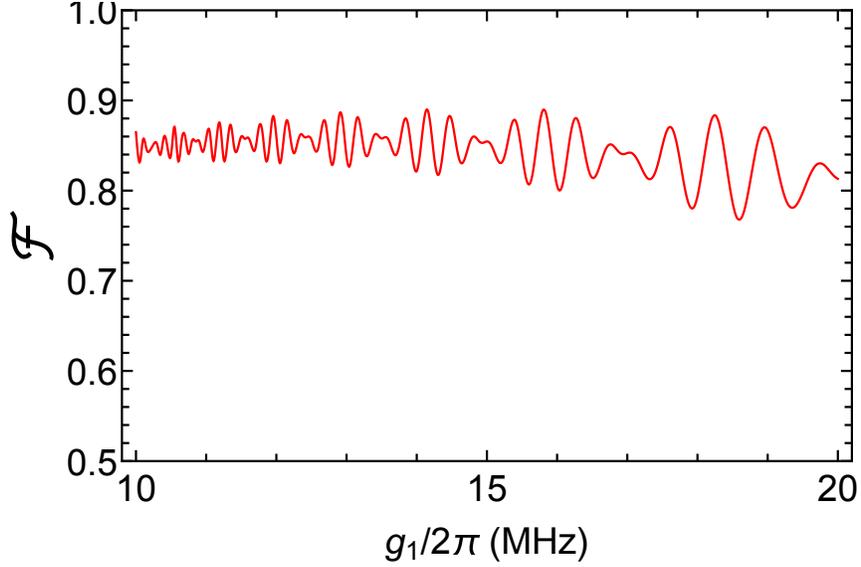} \vspace*{%
-0.08in}
\end{center}
\caption{(color online) Fidelity versus $g_{1}$. The parameters used in the
numerical simulation are referred to the text}
\label{fig:4}
\end{figure}

\begin{figure}[tbp]
\begin{center}
\includegraphics[bb=0 0 436 292, width=11.5 cm, clip]{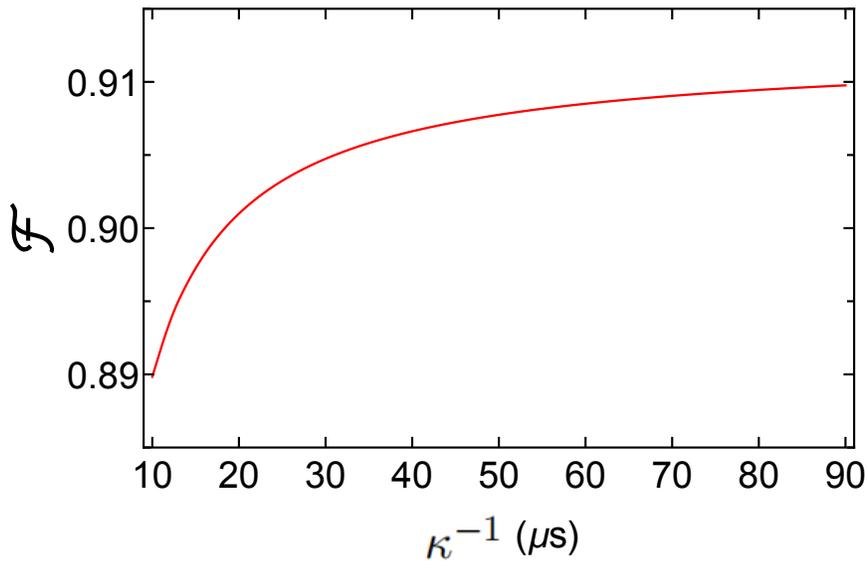} \vspace*{%
-0.08in}
\end{center}
\caption{(color online) Fidelity versus $\protect\kappa ^{-1}$ for $g_{1}/2%
\protect\pi =$ $14.15$ MHz and $\Omega _{l}/2\protect\pi =45$ MHz. Other
parameters used in the numerical simulation are the same as those used in
Fig. 6.}
\label{fig:4}
\end{figure}

To see how the fidelity changes with the cavity decay rate, we plot Fig. 7,
which shows the fidelity versus $\kappa ^{-1}$ for $g_{1}/2\pi =$ $14.15$
MHz and $\Omega _{l}/2\protect\pi =45$ MHz. Fig.~7
demonstrates that the fidelity strongly depends on the photon lifetime of
the cavities. For $\kappa ^{-1}=20$ $\mu $s, a high fidelity $>90\%$ can be
achieved. We remark that the fidelity can be further increased by improving the system parameters.

The operation time is $\sim 0.27$ $\mu $s, which is much shorter than the
decoherence times of transmon qutrits used in our numerical simulations. For
a transmon qutrit, the typical transition frequency between two neighbor
levels is $1-20$ GHz. As an example, we consider $\omega _{eg}/2\pi \sim 6.7$
GHz and $\omega _{fe}/2\pi \sim 6.0$ GHz for the case of the transmon
qutrits being dispersively coupled to their cavities. Thus, for the values
of $\Delta _{1},\Delta _{2},\Delta _{3},\Delta _{4}$ chosen above, one has $%
\omega _{c_{1}}/2\pi =\omega _{c_{3}}/2\pi =6.6$ GHz and $\omega
_{c_{2}}/2\pi =\omega _{c_{4}}/2\pi =6.62$ GHz. For the cavity frequencies
here and $\kappa^{-1}=10$ $\mu $s, the quality factors of the four cavities are $%
Q_{1},Q_{3}\sim 4.14\times 10^{5}$ and $Q_{2},Q_{4}\sim 4.16\times 10^{5}$,
which are available because TLRs with a loaded quality factor $Q\sim 10^{6}$ have been
experimentally demonstrated [69,70]. The analysis given above shows that
high-fidelity creation of GHZ states of four-group SC qubits distributed in
four cavities is feasible with the present circuit QED technology.

Further investigation on the experimental feasibility of
creating GHZ states of more qubits distributed in different cavities would
be necessary. However, we note that the numerical
simulations become rather lengthy and complex as the number of qubits
increases, which is beyond the scope of this theoretical work.

\begin{center}
\textbf{V. CONCLUSION}
\end{center}

We have presented an approach to generate Greenberger-Horne-Zeilinger (GHZ)
entangled states of multiple groups of qubits distributed in multiple
cavities. From the above description, one can see that as long as the
cavities are initially prepared in a GHZ state, all qubits in the cavities
can be entangled via a 3-step operation only, no matter what type of
architecture the cavity-based quantum network preserves and in which way the cavities are
coupled. This proposal also has some additional advantages stated in the
introduction. Our numerical simulation shows that high-fidelity preparation
of GHZ states of four-group SC qubits, each group containing three qubits
and the four groups distributed in four cavities, is feasible with current
circuit QED technology. By increasing the number of resonators,
GHZ states of more groups of SC qubits distributed in multiple cavities can be
created. This work opens a way for quantum state engineering with many qubits
distributed in different cavity nodes of a quantum network. We wish that
it will stimulate experimental activities in the near future.

As a final note, it should be stressed that this proposal is based on the
prerequisite that the cavities are initially prepared in a GHZ state.
Nevertheless, this work is of interest, because it may be easy to entangle
the cavities when compared to directly entangle a large number of qubits
distributed in different cavities without aid of the cavity initial GHZ
states and because the proposal works for a 1D, 2D, or 3D quantum network
composed of cavities.

\begin{center}
\textbf{ACKNOWLEDGMENTS}
\end{center}

This work was partly supported by the Key R$\&$D Program of Guangdong province
(2018B030326001), the National Natural Science Foundation of
China (NSFC) (11074062, 11374083, 11774076), the NKRDP of China
(2016YFA0301802), and the Jiangxi Natural Science Foundation
(20192ACBL20051).

\textbf{APPENDIX: PREPARATION OF THE GHZ STATE OF THE FOUR TLRs}

The ladder-type three levels of each of the coupler qutrits ($%
q_{1},q_{2},q_{3}$) in Fig. 4 are labeled as $\left\vert g\right\rangle ,$ $%
\left\vert e\right\rangle ,$ and $\left\vert f\right\rangle $ with energy $%
E_{g}<E_{e}<E_{f}.$ Initially, $q_{1}$ is in the state $\left(
\left\vert e\right\rangle +\left\vert f\right\rangle \right) /\sqrt{2},$ $%
q_{2}$ and $q_{3}$ are in the ground state $\left\vert
g\right\rangle ,$ and each TLR is in a vacuum state. In
addition, assume that $q_{1},$ $q_{2}$ and $q_{3}$ are decoupled
from their neighbor TLRs. Previously, we have set $\omega _{c_{1}}=\omega _{c_{3}}$ and $\omega _{c_{2}}=\omega
_{c_{4}}$ in Fig. 4, i.e., every two neighbor TLRs have different frequencies.

The procedure for preparing the GHZ state $\left( \left\vert 0\right\rangle
_{c_{1}}\left\vert 0\right\rangle _{c_{2}}\left\vert 1\right\rangle
_{c_{3}}\left\vert 1\right\rangle _{c_{4}}+\left\vert 1\right\rangle
_{c_{1}}\left\vert 1\right\rangle _{c_{2}}\left\vert 0\right\rangle
_{c_{3}}\left\vert 0\right\rangle _{c_{4}}\right) /\sqrt{2}$ of the four
TLRs is listed as follows:

Step 1: Adjust the level spacings of $q_{2}$ such that TLR $2$ is
resonant with the $\left\vert g\right\rangle \leftrightarrow $ $\left\vert
e\right\rangle $ transition of $q_{2},$ with a coupling constant $\mu _{1}.$
After an interaction time $\pi /\left( 2\mu _{1}\right) $ (i.e., half a Rabi
oscillation), the state $\left\vert e\right\rangle _{q_{2}}\left\vert
0\right\rangle _{c_{2}}$ changes to $-i\left\vert g\right\rangle
_{q_{2}}\left\vert 1\right\rangle _{c_{2}}.$ Hence, the initial state $\frac{%
1}{\sqrt{2}}\left( \left\vert e\right\rangle _{q_{2}}+\left\vert
f\right\rangle _{q_{2}}\right) \left\vert 0\right\rangle _{c_{2}}\left\vert
0\right\rangle _{c_{3}}$ of the system, composed of ($q_{2} $, TLR $2$
and TLR $3$), becomes
\begin{equation}
\frac{1}{\sqrt{2}}\left( -i\left\vert g\right\rangle _{q_{2}}\left\vert
1\right\rangle _{c_{2}}+\left\vert f\right\rangle _{q_{2}}\left\vert
0\right\rangle _{c_{2}}\right) \left\vert 0\right\rangle _{c_{3}}.
\end{equation}%
({\it In the following, the normalization factor $\frac{1}{\sqrt{2}}$ will be omitted for simplicity}). Then, adjust the level spacings of $q_{2}$ such that $q_{2}$ is decoupled
from TLR $2.$ Now apply a classical pulse (resonant with the $\left\vert g\right\rangle \leftrightarrow $ $%
\left\vert e\right\rangle $ transition) to $q_{2}$ to pump the state $%
\left\vert g\right\rangle $ back to the state $\left\vert e\right\rangle $.
Thus, the state (23) changes to

\begin{equation}
\left( -i\left\vert e\right\rangle _{q_{2}}\left\vert
1\right\rangle _{c_{2}}+\left\vert f\right\rangle _{q_{2}}\left\vert
0\right\rangle _{c_{2}}\right) \left\vert 0\right\rangle _{c_{3}}.
\end{equation}

Step 2: Adjust the level spacings of $q_{2}$ such that TLR $2$ is
resonant with the $\left\vert g\right\rangle \leftrightarrow $ $\left\vert
e\right\rangle $ transition of $q_{2}$ again. After an interaction time $\pi
/\left( 2\sqrt{2}\mu _{1}\right) $, we have the transformation $\left\vert
e\right\rangle _{q_{2}}\left\vert 1\right\rangle _{c_{2}}$ $\rightarrow $ $%
-i\left\vert g\right\rangle _{q_{2}}\left\vert 2\right\rangle _{c_{2}}$
while the state $\left\vert f\right\rangle _{q_{2}}\left\vert 0\right\rangle
_{c_{2}}$ remains unchanged. Hence, the state (24) becomes
\begin{equation}
\left(-\left\vert g\right\rangle _{q_{2}}\left\vert
2\right\rangle _{c_{2}}+\left\vert f\right\rangle _{q_{2}}\left\vert
0\right\rangle _{c_{2}}\right) \left\vert 0\right\rangle _{c_{3}}.
\end{equation}%
Then, adjust the level spacings of $q_{2}$ such that $q_{2}$ is decoupled
from TLR $2.$

Step 3: Adjust the level spacings of $q_{2}$ such that TLR $3$ is
resonant with the $\left\vert e\right\rangle \leftrightarrow $ $\left\vert
f\right\rangle $ transition of $q_{2},$ with a coupling constant $\mu _{2}.$
After an interaction time $\pi /\left( 2\mu _{2}\right) $, the state $%
\left\vert f\right\rangle _{q_{2}}\left\vert 0\right\rangle _{c_{3}}$
changes to $-i\left\vert e\right\rangle _{q_{2}}\left\vert 1\right\rangle
_{c_{3}}.$ Thus, the state (25) becomes
\begin{equation}
\left\vert g\right\rangle _{q_{2}}\left\vert
2\right\rangle _{c_{2}}\left\vert 0\right\rangle _{c_{3}}+i\left\vert
e\right\rangle _{q_{2}}\left\vert 0\right\rangle _{c_{2}}\left\vert
1\right\rangle _{c_{3}}.
\end{equation}%
Then, adjust the level spacings of $q_{2}$ such that $q_{2}$ is decoupled
from TLR $3.$ Now apply a classical pulse (resonant with the $\left\vert e\right\rangle \leftrightarrow
\left\vert f\right\rangle $ transition) to $q_{2}$ to pump the state
$\left\vert e\right\rangle $ back to the state $\left\vert f\right\rangle $.
Thus, the state (26) changes to

\begin{equation}
\left\vert g\right\rangle _{q_{2}}\left\vert
2\right\rangle _{c_{2}}\left\vert 0\right\rangle _{c_{3}}+i\left\vert
f\right\rangle _{q_{2}}\left\vert 0\right\rangle _{c_{2}}\left\vert
1\right\rangle _{c_{3}}.
\end{equation}

Step 4: Apply a classical pulse (resonant with the $\left\vert g\right\rangle \leftrightarrow $ $\left\vert
e\right\rangle $ transition) to $q_{2}$ to pump the state $\left\vert
g\right\rangle $ to the state $\left\vert e\right\rangle $. Thus, the state (27) changes to

\begin{equation}
\left\vert e\right\rangle _{q_{2}}\left\vert
2\right\rangle _{c_{2}}\left\vert 0\right\rangle _{c_{3}}+i\left\vert
f\right\rangle _{q_{2}}\left\vert 0\right\rangle _{c_{2}}\left\vert
1\right\rangle _{c_{3}}.
\end{equation}%
Then, adjust the level spacings of $q_{2}$ such that TLR $3$ is resonant
with the $\left\vert e\right\rangle \leftrightarrow $ $\left\vert
f\right\rangle $ transition of $q_{2}$ again. After an interaction time $\pi
/\left( 2\sqrt{2}\mu _{2}\right) $, one has the transformation $\left\vert
f\right\rangle _{q_{2}}\left\vert 1\right\rangle _{c_{3}}$ $\rightarrow $ $%
-i\left\vert e\right\rangle _{q_{2}}\left\vert 2\right\rangle _{c_{3}}$
while the state $\left\vert e\right\rangle _{q_{2}}\left\vert 0\right\rangle
_{c_{3}}$ remains unchanged. Thus, the state (28) changes to

\begin{equation}
\left( \left\vert 2\right\rangle _{c_{2}}\left\vert
0\right\rangle _{c_{3}}+\left\vert 0\right\rangle _{c_{2}}\left\vert
2\right\rangle _{c_{3}}\right) \left\vert e\right\rangle _{q_{2}}.
\end{equation}%
The, adjust the level spacings of $q_{2}$ such that $q_{2}$ is decoupled
from TLR $3.$

From the description given above, one can see that TLR $2$ is decoupled
from $q_{2}$ during the operation of steps (3) and (4). In addition, it is
noted that the initial states of TLRs \{$1,4\}$ and coupler qutrits \{$%
q_{1},q_{3}\}$ in Fig. 4 remain unchanged because they are not involved
during each operation of steps $(1)-(4)$ above. Thus, based on Eq.~(29), the state of the whole system after the above 4-step operation is
\begin{equation}
\left( \left\vert 2\right\rangle _{c_{2}}\left\vert
0\right\rangle _{c_{3}}+\left\vert 0\right\rangle _{c_{2}}\left\vert
2\right\rangle _{c_{3}}\right) \left\vert e\right\rangle _{q_{2}}\left\vert
g\right\rangle _{q_{1}}\left\vert g\right\rangle _{q_{3}}\left\vert
0\right\rangle _{c_{1}}\left\vert 0\right\rangle _{c_{4}}.
\end{equation}

The purpose of the remaining operations, described below, is to transfer one
photon from TLR $2$ to TLR $1$ via $q_{1}$ and one photon from TLR $%
3$ to TLR $4$ via $q_{3}.$

Step 5: Adjust the level spacings of $q_{1}$ such that TLR $2$ is
resonant with the $\left\vert g\right\rangle \leftrightarrow $ $\left\vert
e\right\rangle $ transition of $q_{1},$with a coupling constant $\mu _{3}$
After an interaction time $\pi /\left( 2\sqrt{2}\mu _{3}\right) $, the state
$\left\vert g\right\rangle _{q_{1}}\left\vert 2\right\rangle _{c_{2}}$ $%
\rightarrow $ $-i\left\vert e\right\rangle _{q_{1}}\left\vert 1\right\rangle
_{c_{2}}$ while the state $\left\vert g\right\rangle _{q_{1}}\left\vert
0\right\rangle _{c_{2}}$ remains unchanged. Thus, the state (30) becomes
\begin{equation}
\left( -i\left\vert 1\right\rangle _{c_{2}}\left\vert
0\right\rangle _{c_{3}}\left\vert e\right\rangle _{q_{1}}+\left\vert
0\right\rangle _{c_{2}}\left\vert 2\right\rangle _{c_{3}}\left\vert
g\right\rangle _{q_{1}}\right) \left\vert e\right\rangle _{q_{2}}\left\vert
g\right\rangle _{q_{3}}\left\vert 0\right\rangle _{c_{1}}\left\vert
0\right\rangle _{c_{4}}.
\end{equation}%
Then, adjust the level spacings of $q_{1}$ such that TLR $2$ is decoupled
from $q_{1}$ but TLR $1$ is resonant with the $\left\vert g\right\rangle
\leftrightarrow $ $\left\vert e\right\rangle $ transition of $q_{1},$ with a
coupling constant $\mu _{4}.$ After an interaction time $\pi /\left( 2\mu
_{4}\right) ,$ we have the transformation $\left\vert e\right\rangle
_{q_{1}}\left\vert 0\right\rangle _{c_{1}}$ $\rightarrow $ $-i\left\vert
g\right\rangle _{q_{1}}\left\vert 1\right\rangle _{c_{1}}$ while the state $%
\left\vert g\right\rangle _{q_{1}}\left\vert 0\right\rangle _{c_{1}}$
remains unchanged. Hence, the state (31) changes to
\begin{equation}
\left( -\left\vert 1\right\rangle _{c_{1}}\left\vert
1\right\rangle _{c_{2}}\left\vert 0\right\rangle _{c_{3}}+\left\vert
0\right\rangle _{c_{1}}\left\vert 0\right\rangle _{c_{2}}\left\vert
2\right\rangle _{c_{3}}\right) \left\vert g\right\rangle _{q_{1}}\left\vert
e\right\rangle _{q_{2}}\left\vert g\right\rangle _{q_{3}}\left\vert
0\right\rangle _{c_{4}}.
\end{equation}
Then, adjust the level spacings of $q_{1}$ such that both TLRs $1$ and $%
2 $ are decoupled from $q_{1}.$

Step 6: Adjust the level spacings of $q_{3}$ such that TLR $3$ is
resonant with the $\left\vert g\right\rangle \leftrightarrow $ $\left\vert
e\right\rangle $ transition of $q_{3},$ with a coupling constant $\mu _{5}$.
After an interaction time $\pi /\left( 2\sqrt{2}\mu _{5}\right) $, the state
$\left\vert g\right\rangle _{q_{3}}\left\vert 2\right\rangle _{c_{3}}$ $%
\rightarrow $ $-i\left\vert e\right\rangle _{q_{3}}\left\vert 1\right\rangle
_{c_{3}}$ while the state $\left\vert g\right\rangle _{q_{3}}\left\vert
0\right\rangle _{c_{3}}$ remains unchanged. Thus, the state (32) becomes

\begin{equation}
\left(\left\vert 1\right\rangle _{c_{1}}\left\vert
1\right\rangle _{c_{2}}\left\vert 0\right\rangle _{c_{3}}\left\vert
g\right\rangle _{q_{3}}+i\left\vert 0\right\rangle _{c_{1}}\left\vert
0\right\rangle _{c_{2}}\left\vert 1\right\rangle _{c_{3}}\left\vert
e\right\rangle _{q_{3}}\right) \left\vert g\right\rangle _{q_{1}}\left\vert
e\right\rangle _{q_{2}}\left\vert 0\right\rangle _{c_{4}}.
\end{equation}%
Then, adjust the level spacings of $q_{3}$ such that TLR $3$ is decoupled
from $q_{3}$ but TLR $4$ is resonant with the $\left\vert g\right\rangle
\leftrightarrow $ $\left\vert e\right\rangle $ transition of $q_{3},$ with a
coupling constant $\mu _{6}.$ After an interaction time $\pi /\left( 2\mu
_{6}\right) ,$ we have the transformation $\left\vert e\right\rangle
_{q_{3}}\left\vert 0\right\rangle _{c_{4}}$ $\rightarrow $ $-i\left\vert
g\right\rangle _{q_{3}}\left\vert 1\right\rangle _{c_{4}}$ while the state $%
\left\vert g\right\rangle _{q_{3}}\left\vert 0\right\rangle _{c_{4}}$
remains unchanged. Therefore, the state (33) becomes
\begin{equation}
\left( \left\vert 1\right\rangle _{c_{1}}\left\vert
1\right\rangle _{c_{2}}\left\vert 0\right\rangle _{c_{3}}\left\vert
0\right\rangle _{c_{4}}+\left\vert 0\right\rangle _{c_{1}}\left\vert
0\right\rangle _{c_{2}}\left\vert 1\right\rangle _{c_{3}}\left\vert
1\right\rangle _{c_{4}}\right) \left\vert g\right\rangle _{q_{1}}\left\vert
e\right\rangle _{q_{2}}\left\vert g\right\rangle _{q_{3}}.
\end{equation}
Then, adjust the level spacings of $q_{3}$ such that both TLRs $3$ and $%
4 $ are decoupled from $q_{3}.$ Eq.~(34) shows that the four
TLRs are prepared in the GHZ state $\left( \left\vert 0\right\rangle
_{c_{1}}\left\vert 0\right\rangle _{c_{2}}\left\vert 1\right\rangle
_{c_{3}}\left\vert 1\right\rangle _{c_{4}}+\left\vert 1\right\rangle
_{c_{1}}\left\vert 1\right\rangle _{c_{2}}\left\vert 0\right\rangle
_{c_{3}}\left\vert 0\right\rangle _{c_{4}}\right) /\sqrt{2},$ while the
three coupler qutrits ($q_{1},q_{2},q_{3}$) are disentangled from the four
TLRs.

Since each step of operation employs the resonant qutrit-cavity or
qutrit-pulse interaction, the GHZ state of the four TLRs can be fast
prepared within a short time.


\begin{thebibliography}{99}
\bibitem{s64} For a transmon qutrit, the $\left\vert 0\right\rangle
\leftrightarrow \left\vert 2\right\rangle $ transition is much weaker than
those of the $\left\vert 0\right\rangle \leftrightarrow \left\vert
1\right\rangle $ and $\left\vert 1\right\rangle \leftrightarrow \left\vert
2\right\rangle $ transitions. Thus, we have $\gamma _{20}^{-1}\gg \gamma
_{10}^{-1},\gamma _{21}^{-1}$.
\end{thebibliography}
\end{document}